\documentclass{jpp}
\usepackage{graphicx}
\usepackage[utf8]{inputenc}
\usepackage[T1]{fontenc}
\usepackage{amsmath}
\usepackage{float}
\usepackage{hyperref}
\usepackage[dvipsnames]{xcolor}
\usepackage{siunitx}

\newcommand{\dream}{\textsc{Dream}}
\newcommand{\dBB}{\delta B/B}
\newcommand{\Epa}{E_\|}
\newcommand{\Eceff}{E_{\rm c}^{\rm eff}}
\newcommand{\ndi}{n_{\rm D,inj}}
\newcommand{\nnei}{n_{\rm Ne,inj}}

\newcommand{\xv}{\mathbf{x}}

\shorttitle{Optimization of massive material injection}
\shortauthor{I.~Pusztai et al}

\title{Bayesian optimization of massive material injection for disruption mitigation in tokamaks}

\author{I.~Pusztai\aff{1}
  \corresp{\email{pusztai@chalmers.se}},  I.~Ekmark\aff{1}, H.~Bergstr\"{o}m\aff{2,1}, P.~Halldestam\aff{2,1}, P.~Jansson\aff{3}, M.~Hoppe\aff{4}, O.~Vallhagen\aff{1},
   \and T.~Fülöp\aff{1}}

\affiliation{\aff{1}Department of Physics, Chalmers University of Technology, G\"{o}teborg, SE-41296, Sweden
\aff{2} Max Planck Institute for Plasma Physics, Garching b. M, 85748, Germany
\aff{3} Department of Computer Science and Engineering, Chalmers University of Technology, Göteborg, SE-41296, Sweden
\aff{4}Swiss Plasma Center, Ecole Polytechnique F\'{e}d\'{e}rale de Lausanne, Lausanne, CH-1015, Switzerland}

\begin{document}

\maketitle

\begin{abstract}
A  Bayesian optimization framework is used to investigate scenarios for disruptions mitigated with combined deuterium and neon injection in ITER. The optimization cost function takes into account limits on the maximum runaway current, the transported fraction of the heat loss and the current quench time. The aim is to explore the dependence of the cost function on injected densities, and provide insights into the behaviour of the disruption dynamics for representative scenarios. The simulations are conducted using the numerical framework \dream{} (Disruption Runaway Electron Analysis Model). We show that irrespective of the quantities of the material deposition, multi-megaampere runaway currents will be produced in the deuterium-tritium phase of operations, even in the optimal scenarios. However, the severity of the outcome can be influenced by tailoring the radial profile of the injected material; in particular if the injected neon is deposited at the edge region it leads to a significant reduction of both the final runaway current and the transported heat losses. The Bayesian approach allows us to map the parameter space efficiently, with more accuracy in favorable parameter regions, thereby providing us information about the robustness of the optima.

\end{abstract}

\section{Introduction}
\label{sec:intro}
One of the threats to reliable tokamak operation are off-normal
events known as disruptions, which are induced by a sudden loss of plasma confinement \citep{BoozerDisr}. When this occurs, the ensuing heat and particle transport results in a rapid temperature drop -- a thermal quench (TQ) -- that is accompanied by a decrease in the electrical conductivity of the plasma. The reduced conductivity leads to a decay in plasma current -- a current quench (CQ) --  that is counteracted by
the induction of an electric field, which may accelerate runaway electrons (REs) to relativistic energies \citep{Breizman_2019}. The REs could potentially strike the wall and lead to subsurface melting of the wall components. 

The plasma current in future devices will be around an order of magnitude higher than in present experiments.  Correspondingly, the magnetic energy in the plasma will increase ($\sim \SI{400}{MJ}$ in ITER  versus $\sim \SI{10}{MJ}$ in JET) \citep{Hender_2007}, along with the kinetic energy, thus the available energy that can be released in a disruption is significantly higher than in present devices. It is therefore essential to develop effective disruption mitigation systems.

An effective disruption mitigation system in a tokamak should limit the exposure of the wall to localized heat losses and to the impact of high current RE beams, and avoid excessive forces on the structure \citep{Hollmann}.   To avoid damage to the first wall on ITER, at least
$90\%$ of the thermal energy loss must be lost in the form of radiation.  The RE current should be kept below $150\,\rm kA$ in order to avoid melting of plasma facing components, in the case of localised loss \citep{Lehnen_talk}. The CQ time, i.e.~the time it takes for the ohmic component of the current to decay, should be kept between $50$ and $150\,\rm ms$. Current quench times below $50\,\rm ms$ will lead to excessive forces due to eddy currents in the structures surrounding the plasma. On the other hand, CQ times above $150\,\rm ms$ are expected to lead to intolerably large halo currents in plasma facing components. 

In ITER, the envisaged disruption mitigation system is based on massive material injection \citep{Lehnen2015}. The injected material can radiate away a large fraction of the thermal energy and it can also inhibit RE generation by increasing the critical energy for electron runaway. Furthermore, it can also be used to control the temperature during the CQ, which directly influences the CQ duration. However, the question of what mixture of material should be injected, and how it should be deposited, to accommodate all requirements on the disruption mitigation system simultaneously, if it is at all possible, is still open.

In this paper, we describe a Bayesian optimization framework applied to simulations of ITER-like disruption scenarios mitigated with combined injection of deuterium and neon. The aim is to find the injected material quantities and deposition profiles for which the outcome of the disruption is tolerable with respect to the expected RE current, transported heat fraction and CQ time. Bayesian optimization has several attractive features: it does not rely on gradient information, it can handle non-deterministic (noisy) functions, and it is suitable for relatively high-dimensional optimization problems and computationally expensive function evaluations. However, its main advantage concerning the current study is that it informs us about the properties of promising parameter regions -- in particular the robustness of the optima to variations in the control parameters. 

The rest of the paper is structured as follows. The methods are explained in Sec.~\ref{sec:Methods}, detailing the setup of the disruption simulations in \ref{sec:setup} and the Bayesian optimization in \ref{sec:optim}. The results are presented in Sec.~\ref{sec:results}, first mapping out the optimization landscape with constant injected densities in \ref{sec:ladscape} followed by a detailed analysis of some representative scenarios in \ref{sec:cases}. Then we present optimization results  allowing radially varying injection in \ref{sec:radvar}. Finally we study the parametric sensitivities of the optima and reflect on the beneficial effects of radial profile variations in \ref{sec:sensitivity}, before we conclude and discuss our findings in Sec.~\ref{sec:conclusions}.       

\section{Bayesian optimization of simulated disruptions}
\label{sec:Methods}
We employ an open source Bayesian optimization routine that treats the disruption simulations as a black-box function that produces a single scalar output, {\em the cost function}, and accepts inputs for injected material densities and deposition profiles in specified ranges -- these are the input parameters that we want to optimize. In the following we will discuss the disruption simulations, and provide details of the optimization algorithm.    

\subsection{Simulation setup}
\label{sec:setup}

The disruption simulations assume an initially ($t<0$) pure fully-ionized deuterium-tritium (D-T) plasma with $50$--$50\%$ isotope concentrations. 
Specifically, the initial electron density is spatially constant $10^{20}\,\rm m^{-3}$, the temperature is parabolic with $20\,\rm keV$ on-axis, and the  total plasma current is $15\,\rm MA$. The simulations use an ITER-like magnetic geometry with major radius $R_0=6\,\rm m$, minor radius $a=2\,\rm m$, wall radius $b=2.833\, \rm m$,  on-axis toroidal magnetic field $B(r=0)=5.3\,\rm T$, and a resistive wall time of $\tau_w=0.5\,\rm s$, as well as a Miller model equilibrium \citep{Miller98} with  realistic, radially varying shaping parameters; further information given in Appendix~\ref{sec:simulations}.

The simulations are performed by the \dream{} (Disruption Runaway Electron Analysis Model) code that captures the particle acceleration and energy dissipation processes following a disruption \citep{dreampaper}. It solves a set of coupled transport equations describing the evolution of temperature, ion charge state densities, current density and electric field in arbitrary axisymmetric geometry. The temperature evolution includes ohmic heating, radiated power using atomic rate coefficients, collisional energy transfer from hot electrons and ions, as well as dilution cooling. 

\dream{} allows modelling of the REs at different degrees of approximation ranging from fluid to fully kinetic. As we do not require kinetic outputs,  we limit our modelling to the least computationally expensive, fluid treatment of the plasma. This means that the thermal bulk of cold electrons and the small runaway population are modelled as two separate fluid species. The former is characterized by a density $n_e$, a temperature $T_e$ as well as an ohmic current density $j_{\rm ohm}$, and the REs are described by their density
$n_{\rm RE}$. It is assumed that the REs move with the speed of light parallel to the
magnetic field, hence their associated current density is $j_{\rm RE} = ecn_{\rm RE}$. The simulations include Dreicer, hot-tail, and avalanche sources, as well as REs generated by Compton scattering of $\gamma$ photons and tritium decay. These are modelled as quasi-stationary sources feeding electrons into the runaway population \citep{Fulop20}. The runaway generation rates used in the simulations have been benchmarked with the corresponding  kinetic results \citep{dreampaper}. Further details on the simulations are given in Appendix~\ref{sec:simulations}. 

Neutral neon and deuterium are introduced with zero temperature at the start of the simulation ($t=0$).  At the same time an elevated transport of electron heat and energetic electrons is activated, using a Rechester-Rosenbluth-type model \citep{RechRos} with a radially constant normalized magnetic perturbation amplitude $\dBB$.  This is done to emulate the break-up of flux surfaces during the TQ, and leads to heat-losses, with a heat diffusivity proportional to $R_0 v_{te} (\dBB)^2$, where  $v_{te}=\sqrt{2 T_e/m_e}$ is the local electron thermal speed. The full expression is given by Eq.~(B.5) of \citep{dreampaper}. In four optimization runs $\dBB$ is scanned over the range $0.2\%-0.5\%$ that falls within the range of values observed in magnetohydrodynamic simulations of the TQ \citep{Hu_2021}. 

During the TQ we also account for a diffusive transport of REs using a diffusion coefficient of similar form, but assuming a parallel streaming along the perturbed field lines at the speed of light, $D_{\rm RE}=\pi R_0 c (\dBB)^2$.  This approach neglects the momentum space variation of the transport coefficients \citep{KonstaNF2020}, as well as the form of the RE distribution function, which would reduce the effect of runaway transport. Thus, using this expression provides an upper bound on the effect of runaway transport for a given magnetic perturbation amplitude \citep{Svensson2021}. We employ here the same $\dBB$ as for electron heat transport for consistency. 

 The injected material is ionized by its interaction with the plasma, and cools it by radiation and dilution. When the average electron temperature falls below $10^{-3}$ times the maximum initial temperature (here $20\,\rm eV$), we assume that the TQ is completed and the flux surfaces reform. After the TQ the transport of energetic electrons is switched off, and a significantly reduced, but finite electron heat diffusivity is used ($\delta B/B=0.04\%$). This is to avoid the development of non-physical narrow hot ohmic channels during the CQ. Such ohmic channels are soliton-like solutions of the problem \citep{Putvinski97} without sufficient heat diffusivity. In a physical system the corresponding excessive temperature and current gradients would be expected to destabilize these formations well before they could fully form. Note, that the diffusive heat transport is subdominant compared to radiative heat losses at the low post-TQ temperatures, thus this heat transport has no effect besides not allowing hot channels to form.

\subsection{Optimization}
\label{sec:optim}
 
The optimization problem involves multiple objectives, i.e.~multiple quantities need to be within certain limits simultaneously. The maximum value of the total RE current  and the fraction of transported heat losses  must be small, while the CQ time  should be within certain limits. These quantities are normalized and combined into a single scalar \emph{cost function} $\mathcal{L} \ge 0$ which is to be minimized. Denoting the control vector containing the parameters by $\xv$, we wish to find the $\xv^{\ast}$ that minimizes $\mathcal{L}$, where $\xv$ resides in a specified volume $\mathcal{V}\subset \mathbb{R}^d$ of the control space (where $d$ is the dimensionality of the optimization).  

We employ Bayesian optimization \citep{bayesian} using Gaussian process regression \citep{Rasmussen2005}, using the \emph{Bayesian Optimization} \citep{bayesopt} Python package. 
A Gaussian process is fitted to the already sampled points $\{\xv_i\}_{i=1}^n$, and the \emph{Expected improvement} acquisition strategy (described in Appendix~\ref{sec:bayesapp}) is used to choose the next point to be sampled, $\xv_{n+1}$.
The Gaussian process contains information on both the expected value $\mu(\xv)$ and the uncertainty of the estimate of $\mathcal{L}$, quantified in terms of the covariance $k(\xv, \xv')$ between any two points $\xv$ and $\xv'$. 
In this process there is a balance between \emph{exploration} and \emph{exploitation}, i.e.~search within regions with high uncertainty, as well as in regions that are most likely to host the global optimum. 

The cost function we use is of the form
\begin{equation}
\mathcal{L}=\frac{I_{\rm RE}^{\rm max}}{I_{\rm RE}^{\rm tol}} + \frac{I_{\rm ohm}^{\rm fin}}{I_{\rm ohm}^{\rm tol}} + 10 \frac{\eta_{\rm cond}}{\eta_{\rm cond}^{\rm tol}} + 100\,\theta(t_{\rm CQ}) ,
    \label{eq:cost}
\end{equation}
where $I_{\rm RE}^{\rm max}$ is the maximum RE current in the simulation, $I_{\rm RE}^{\rm tol}=150\,\rm kA$ represents the tolerable RE current in ITER, $I_{\rm ohm}^{\rm fin}$ is the ohmic current at the end of the ($150\,\rm ms$ long) simulation. A significant remnant ohmic current may be the sign of an incomplete TQ, and it can still potentially be converted to a RE current. Thus it is treated on equal footing with the RE current, so we also set $I_{\rm ohm}^{\rm tol}=150\,\rm kA$. $\eta_{\rm cond}^{\rm tol}=0.1$ is the tolerable transported heat loss fraction. The prefactor $10$ in the $\eta_{\rm cond}$ term is used to get a  penalty for non-tolerable transported heat losses comparable to typical penalties obtained for mega-ampere ($\rm MA$) size currents. Finally, to penalize CQ times $t_{\rm CQ}$ below $t_{\rm L}=50\,\rm ms$ and above $t_{\rm U}=150\,\rm ms$  we use the penalty function
\begin{equation}
    \theta(t_{\rm CQ})=  \tilde{\Theta}(t_{\rm L} - t_{\rm CQ}) + \tilde{\Theta}(t_{\rm CQ} - t_{\rm U}),
\end{equation}
where $\tilde{\Theta}(t)= \frac{1}{2}[1+\tanh (t/\Delta t)]$ is a function similar to a step function but smooth with a transition width set by $\Delta t=3.3 \;\rm ms$. Values of $t_{\rm CQ}$ outside the tolerable range yield a penalty as high as the maximum achievable penalty for any of the other terms in  (\ref{eq:cost}), due to the prefactor $100$ in front of $\theta$. We calculate the CQ time as $t_{\rm CQ}=[t(I_{\rm ohm}=0.2 I_p^{0})-t(I_{\rm ohm}=0.8 I_p^{0})]/0.6$ \citep{Hender_2007}, where $I_{\rm ohm}(t)$ is the total ohmic current and $I_p^{0}$ is the initial plasma current. 

In addition we set $\mathcal{L}=500$ for simulations where the TQ is not complete within $20\,\rm ms$, our condition for which is that the average temperature falls to below $10^{-3} T_e(r=0,t=0)$. Finally, as it is difficult to completely avoid simulations that fail due to numerical issues, we use $\mathcal{L}=500$ for these as well.  

Independently of their dimensionality, the optimizations use $400$ samples, chosen according to the acquisition function through sequential function evaluations, following $20$ randomly selected initial samples. As the parameter space of injected quantities ranges across multiple orders of magnitude, the logarithm of the injected quantities is used as optimization parameters.  

\section{Bayesian optimization of disruption mitigation with material injection}
\label{sec:results}
The goal is to identify what densities of injected neon and deuterium produce the
most favourable outcomes in a disruption mitigation, corresponding to the minimum of the cost function. Modelling the details of the material
injection is outside the scope of the present work, instead we assume the material to be instantaneously deposited in the
form of neutrals, either uniformly distributed over the magnetic flux surfaces, described in Sec.~\ref{sec:ladscape}-\ref{sec:cases} or with radially varying distribution, described in Sec.~\ref{sec:radvar}. 

\subsection{Optimization landscape with constant concentrations}
\label{sec:ladscape}

First we perform optimization in the two-dimensional (2D) parameter space of radially constant injected deuterium and neon densities, $n_{\rm inj,D}$ and $n_{\rm inj,Ne}$.  The ranges of injected densities we consider are $n_{\rm inj,D} \in [10^{18},\,3.16\times 10^{22}]\,\rm m^{-3}$, and $n_{\rm inj,Ne} \in [10^{16},\,10^{20}]\,\rm m^{-3}$. 

\begin{figure}
    \centering
    \includegraphics[width=1\textwidth]{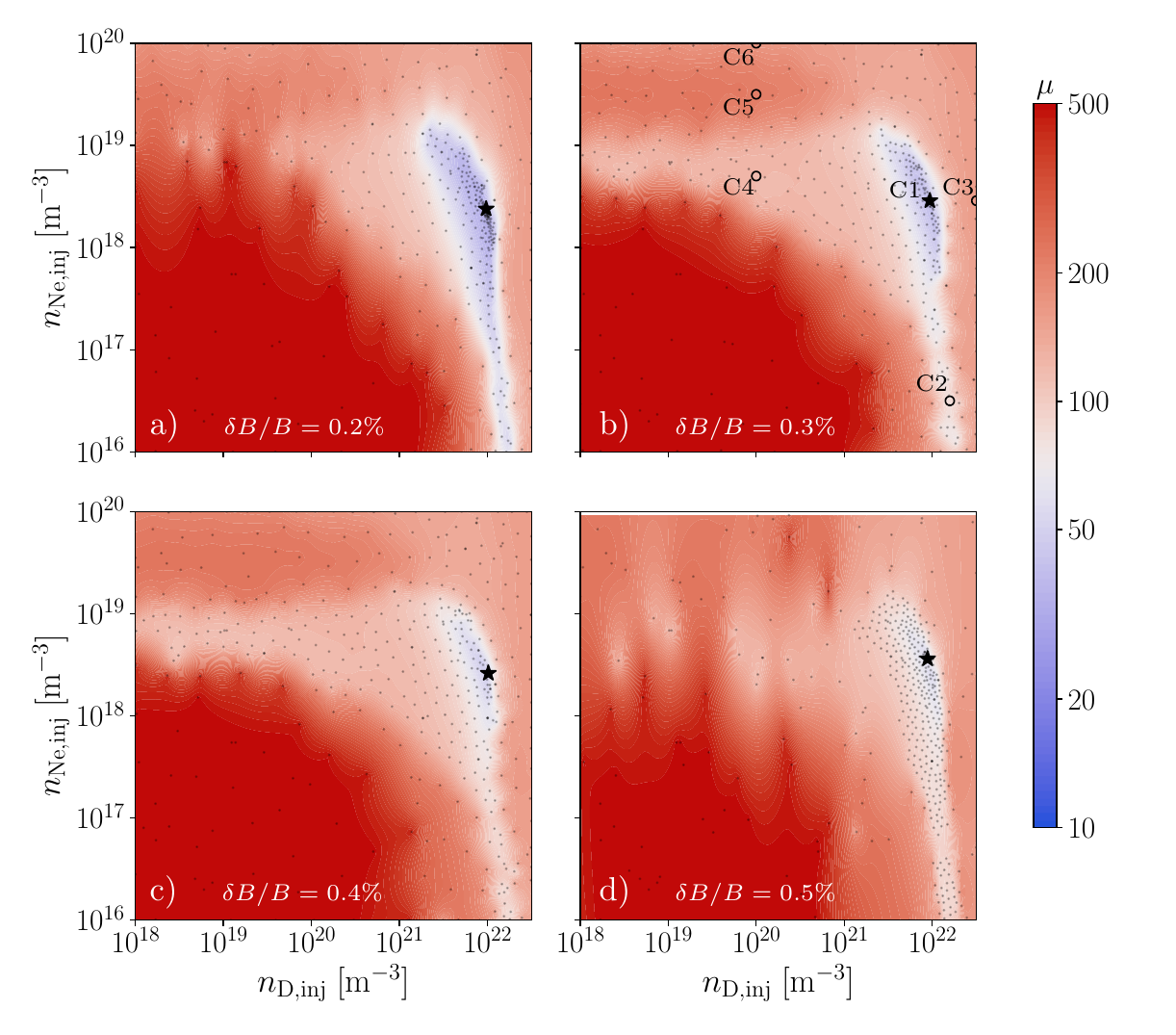}
    \caption{\label{fig:2x2} The estimated mean of the cost function of Bayesian optimizations in the $\ndi$--$\nnei$ space for various normalized magnetic perturbation amplitudes. The color code varies from blue to red tones, representing favourable and unfavourable values of $\mu$. a) $\dBB=0.2\%$, b) $0.3\%$, c) $0.4\%$, d) $0.5\%$. Black stars indicate the locations of the optima. Gray dots show the samples taken; note that these are more numerous in the vicinity of the optima. Circles with case identifiers in panel b) indicate the cases discussed in Sec.~\ref{sec:cases}.}   
\end{figure}

Figure~\ref{fig:2x2} shows the estimated mean of the cost function $\mu$ on a logarithmic contour plot for four different values of $\dBB$, with blue shades representing favourable and red shades unfavourable values. Each subplot used $420$ samples, indicated by gray dots, while the optima are indicated with black stars. The area of favourable values (with blue shades) decreases with  increasing $\dBB$, and this is mostly due to the increasing transported heat fraction, and to a lesser degree to an increasing RE current, to be discussed further in relation to Fig.~\ref{fig:FigsMerit}. In general, the lower left corner of the plots is occupied by cases with an incomplete TQ. In this case the plasma tends to get reheated after the prescribed transport event, leading to long CQ times (i.e.~$t_{\rm CQ}>150\,\rm ms$). With an increasing $\dBB$ the incomplete TQ region shrinks somewhat. 

Another general feature is a relatively narrow corridor of favourable parameters in the vicinity of  $n_{\rm inj,D} = 10^{22}\,\rm m^{-3}$, extending from the lowest $n_{\rm inj,Ne}$ values plotted to a bit above $n_{\rm inj,Ne} = 10^{19}\,\rm m^{-3}$. The optima also reside in these corridors at $n_{\rm inj,Ne}$ values of a few times $10^{18}\,\rm m^{-3}$. Additionally, a wider corridor of moderate values of $\mathcal{L}$ extends to the left of the optima, between $n_{\rm inj,Ne}\approx 3\times 10^{18}$--$2\times 10^{19}\,\rm m^{-3}$, which is most pronounced in the $\dBB=0.3\%$ and $0.4\%$ cases. At $n_{\rm inj,Ne}$ values above that, $\mathcal{L}$ increases, and decreases again around the highest $n_{\rm inj,Ne}$ values included in this optimization. Before analyzing the characteristic behaviors in these different regions of the current optimization landscape in Sec.~\ref{sec:cases}, we shall discuss the detailed dynamics at the optima.

\begin{figure}
    \centering
    \includegraphics[width=0.37\textwidth]{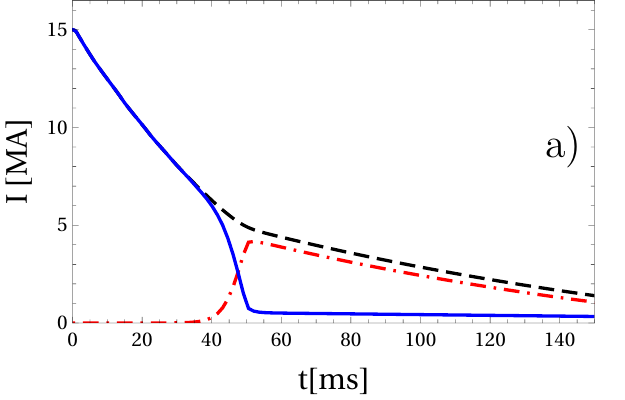}
    \includegraphics[width=0.38\textwidth]{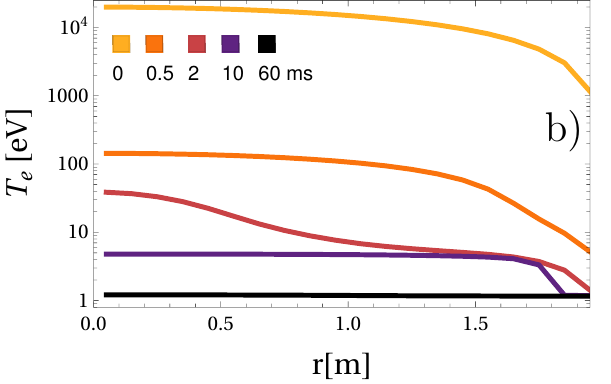}\\
    \includegraphics[width=0.38\textwidth]{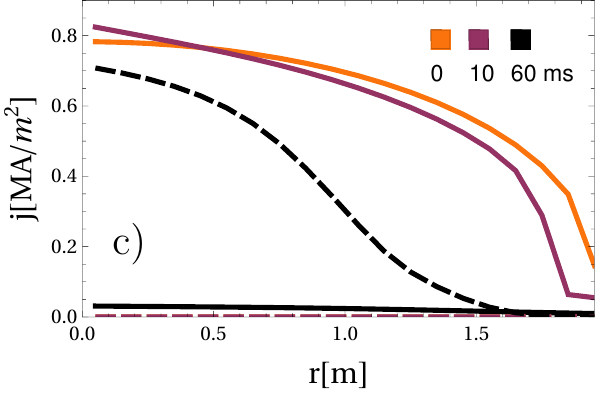}
    \includegraphics[width=0.38\textwidth]{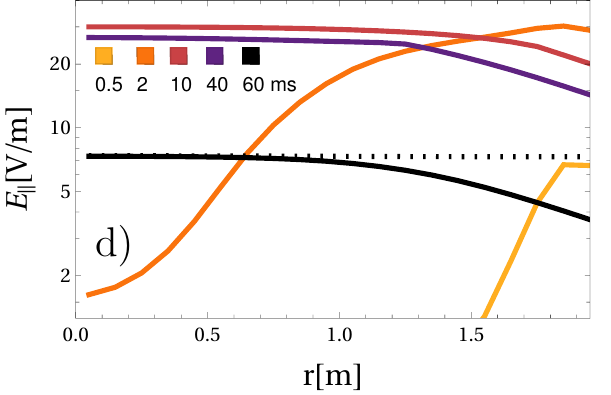}
    \caption{\label{fig:C1Profiles} The best performing case for the optimization in the   $\ndi$--$\nnei$ space, for $\dBB=0.3\%$. a) The time evolution of the total plasma current (dashed), and its ohmic (solid) and RE (dash-dotted) components. b)-d) show radial profiles of quantities in a few time points, indicated by their respective figure legends; with increasing time corresponding to darker colors. b) Electron temperature. c) Ohmic (solid) and RE (dashed) current density. d) Parallel electric field (solid). The effective critical electric field is also indicated for $t=60\,\rm ms$ (dotted); note that it does not vary appreciably over time. } 
\end{figure}

We consider the behavior of the representative optimum obtained in the $\dBB=0.3\%$ case (indicated by the black star in Fig.~\ref{fig:2x2}b), located at $n_{\rm inj,D}=9.4\times 10^{21}\,\rm m^{-3}$ and $n_{\rm inj,Ne}=2.9\times 10^{18}\,\rm m^{-3}$, shown in Fig.~\ref{fig:C1Profiles}. Following the instantaneous material injection at $t=0$, the temperature profile drops by a factor of $\approx 100$ within a $\rm \mu s$, due to dilution. This is followed by an approximately exponential cooling with a characteristic time of $\tau \approx 1.5 \,\rm ms$.
After the initial exponential cooling, a cold front starts to propagate radially inward from the edge. This inward propagating cooling is seen in the $t=2\,\rm ms$ curve in Fig.~\ref{fig:C1Profiles}b. This cooling proceeds until almost the entire plasma settles at around $5\,\rm eV$ (see the $t=10\,\rm ms$ curve), representing the equilibrium between ohmic heating and radiation corresponding to the ion composition and current density of the plasma. Then there is another inward propagating cooling happening over the next $50\,\rm ms$. As the ohmic current density drops during the CQ, the equilibrium temperature falls from $\approx 5\,\rm eV$ to $\approx 1.2\,\rm eV$. The temperature is radially uniform at this level at $60\,\rm ms$ (black curve), and remains there until the end of the simulation.

The ohmic current density gradually decreases in the core and it drops rapidly across the cold front at the edge; compare the $10\,\rm ms$ curves in Fig.~\ref{fig:C1Profiles}b and c. This front propagates inward in the first $40\,\rm ms$, after which the ohmic current gets rapidly replaced by RE current; see the process in terms of total current components in Fig.~\ref{fig:C1Profiles}a, and RE current density at $60\,\rm ms$ in Fig.~\ref{fig:C1Profiles}c (dashed line). The electric field exceeds the effective critical field $\Eceff$ -- calculated as in Appendix C2 of \citep{dreampaper} -- first in the edge, then it grows to an approximately radially constant value around $30 \,\rm V/m$, roughly 4 times $\Eceff$, where it stays until the macroscopic RE conversion starts. Then it drops into the vicinity of $\Eceff$, such that in the core $\Epa$ is  pinned to $\Eceff$, and it takes radially decreasing values at the edge; compare $\Epa$ (solid black curve) to $\Eceff$ (dotted) in Fig.~\ref{fig:C1Profiles}d. Then the electric field remains like that until most of the RE current dissipates away. Physically, the dissipation of the RE current, in the absence of transport losses, is caused by a collisional slowing down and thermalization of the REs when $\Epa<\Eceff$. In the employed fluid RE model it is technically accounted for by allowing the avalanche growth rate to become negative for $\Epa<\Eceff$ values. The corresponding decay of the RE current is quite pronounced in this case.        

\subsection{Characteristic cases with constant concentrations}
\label{sec:cases}

In order to understand the typical dynamics in various regions in the $\ndi$--$\nnei$ space, we consider six representative cases in the $\dBB=0.3\%$ optimization, with case C1 being the optimum discussed above. The cases are indicated in Fig.~\ref{fig:2x2}b and corresponding injected quantities and figures of merit are listed in Tab.~\ref{tab:cases}. Cases C2 and C3 are taken in the high $\ndi$ region of the space; C2 is located in the favourable channel at low $\nnei$, and C3 at even higher $\ndi$ than the optimum. Cases C4 to C6 are taken at a fixed $\ndi=10^{20}\,\rm m^{-3}$, at respectively increasing value of $\nnei$. We discuss C1-C3 and C4-C6 in the following subsections. 

\subsubsection{Representative cases at high $\ndi$}
\label{highndicases}

\begin{table}
  \begin{center}
  \begin{tabular}{cccccccc}
      Case ID & $\ndi[10^{20}m^{-3}]$ & $\nnei\rm [10^{18}m^{-3}]$ & $I_{\rm RE}^{\rm max}\rm [MA]$ & $I_{\rm ohm}^{\rm fin}\rm [MA]$ & $t_{\rm CQ}\rm [ms]$ & $\eta_{\rm cond} [\%]$ & $\mathcal{L}$ \\[3pt]
 C1 & 93.9 & 2.88 & 4.2 & 0.33 & 59 & 8.9 & 39\\ 
 C2 & 160 & 0.032 & 4.8 & 0.33 & 54 & 43 & 88\\
 C3 & 316 & 2.88 & 8.2 & 0.0007 & 5 & 1.4 & 156\\
 C4 & 1 & 5.01 & 6.3 & 0.059  & 88 & 80 & 122\\
 C5 & 1 & 31.6 & 8.1 & 0.092 & 26 & 72 & 226\\
 C6 & 1 & 100 & 8.9 & 0.159 & 15 & 23 & 163\\
  \end{tabular}
  \caption{\label{tab:cases} Characteristic cases from the $\ndi$--$\nnei$ optimization landscape for $\dBB=0.3\%$, their four figures of merit and corresponding cost function values. The cases are marked in Fig.~\ref{fig:2x2}b. C1 is the optimum.}
  \end{center}
\end{table}

\begin{figure}
    \centering
    \includegraphics[width=0.32\textwidth]{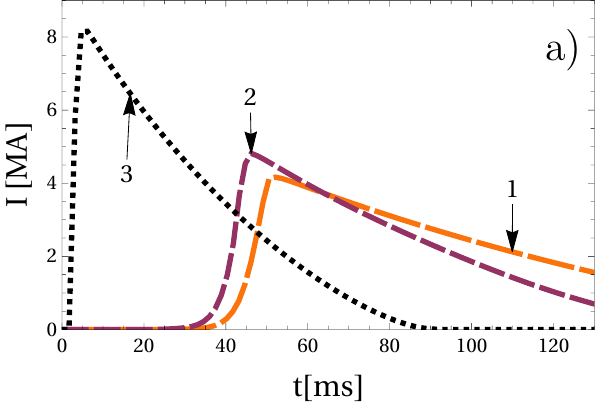}
    \includegraphics[width=0.32\textwidth]{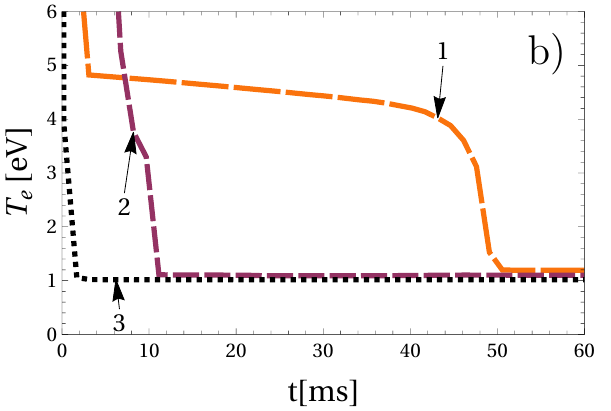}
    \includegraphics[width=0.33\textwidth]{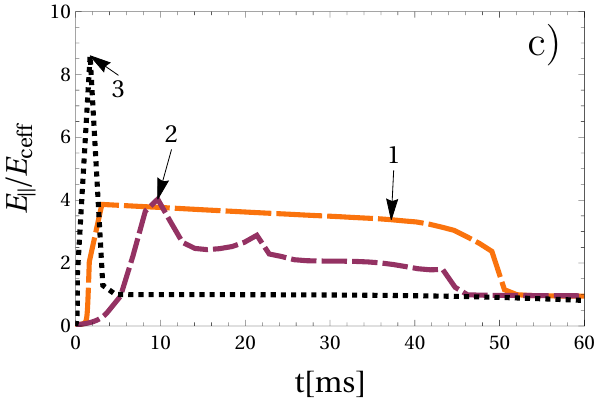}
    \caption{\label{fig:Allcases123}  Time evolution of quantities of interest for the high $\ndi$ representative cases: C1--C3. Line color darkens and dashing shortens with increasing case number, and case numbers are indicated with callouts. a) Runaway electron current. b) Electron temperature at mid-radius. c) Electric field normalized to critical electric field at mid-radius. (Note the longer time range plotted in panel a). }    
\end{figure}

In the RE plateau the electric field tends to stay close to the effective field $\Eceff$, as expected \citep{Breizman_2014}. In particular, following the RE conversion, all $\Epa/\Eceff$ values (taken at mid-radius) settle around unity, as shown in Fig.~\ref{fig:Allcases123}c. The high $\ndi$ cases are all characterized by a significant decay rate of the RE current after it reaches its maximum value; see Fig.~\ref{fig:Allcases123}a. This is consistent with their $\Epa/\Eceff$ being lower than unity towards the edge, as we have seen for C1 in Fig.~\ref{fig:C1Profiles}d. This is due to the relatively high value of $\Eceff$ typical at these high $\ndi$ values.  

The dynamics of the RE current in C1 and C2 are fairly similar, as seen in Fig.~\ref{fig:Allcases123}a. It may be surprising that $\Epa/\Eceff$ is almost all the time higher in C1 than C2 -- shown in Fig.~\ref{fig:Allcases123}c -- but the maximum RE current in C2 is still higher, and is reached a bit earlier. The reason for this is that the temperature drops to $1.09\,\rm eV$ in C2 already at $t\approx 10\,\rm ms$ (Fig.~\ref{fig:Allcases123}b), a temperature where $44\%$ of the hydrogenic species is recombined, thereby increasing the total-to-free electron density ratio and the avalanche growth rate in proportion. Meanwhile in C1 the temperature does not drop to this low temperature  until the RE conversion is over. 

The effect of the hydrogen recombination is even more pronounced in C3, where the temperature drops to $1.02\,\rm eV$ within a millisecond. The reason for this  fast cooling is the very high dilution that brings the temperature down to a range where radiative losses are strong and can effectively (and rapidly) cool the plasma further. That the temperature drops immediately to its final value, without stopping at some higher, intermediate value, can be explained by how the temperature dependence of the total radiative losses ($P$) is affected by the very high hydrogen content. Depending on the hydrogen (including D and T) and neon densities, the curve $P(T_e)$ can exhibit a local minimum in the few $\rm eV$ range between a low $T_e$ peak caused by hydrogen and a higher $T_e$ peak from neon. The large hydrogen density in C3 leads to an elevated value of $P$ at this minimum, thereby effectively eliminating the bottleneck this minimum represents concerning the cooling.   While $1.02\,\rm eV$ is just slightly cooler than the final temperature in C2, now $70\%$ of the hydrogenic species are recombined, which, in combination with the early high value of $\Epa/\Eceff$, leads to an extremely fast RE conversion and the highest RE current among these three cases. 

In terms of figures of merit, C3 is not only problematic due to a high $I_{\rm RE}^{\rm max}$ value, but also because of the extremely short $t_{\rm CQ}\approx 5\,\rm ms$. While $I_{\rm RE}^{\rm max}$ is not too much higher in C2 than in C1, it has a $\eta_{\rm cond}\approx 44\%$, exceeding the tolerable $10\%$, unlike C1 and C3. This is due to the small neon content in C2. 

The remarkably short cooling times, of the order of $2\,\rm ms$, observed at large deuterium injections, such as C3, may be partly due to our simplifying assumption of instantaneous deposition. However, in realistic material injection scenarios, the cooling at a given flux surface can be comparably short to the time observed here,  even if the time-scale needed for pellet shards flying at $500\,\rm m/s$ to travel between the edge and the center of an ITER plasma  is longer ($\approx\,\rm  4 ms$).  As the \emph{local} cooling time is the crucial factor to get a large hot-tail seed, and furthermore, the rapid avalanche rate depends on the final temperature, similar behaviour is also observed in shattered pellet injection simulations \citep{Vallhagen2022}. Ion convection timescales across the radius in a TQ can also be in the $\rm ms$ range. The excessive runaway generation is thus not an artefact of the instantaneous deposition, however, the detailed temperature evolution is expected to be different once the injection dynamics is resolved.    

\subsubsection{Representative cases at low $\ndi$}
\label{lowndicases}

\begin{figure}
    \centering
    \includegraphics[width=0.32\textwidth]{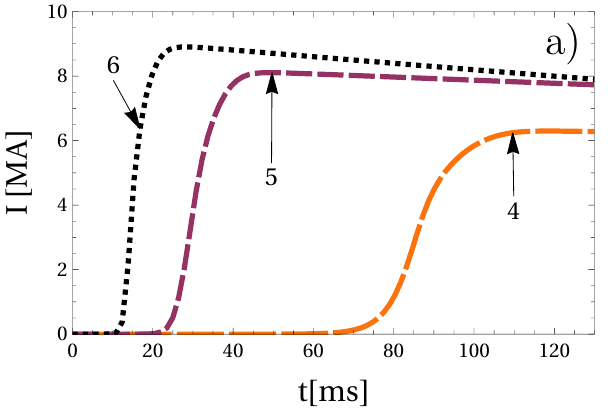}
    \includegraphics[width=0.32\textwidth]{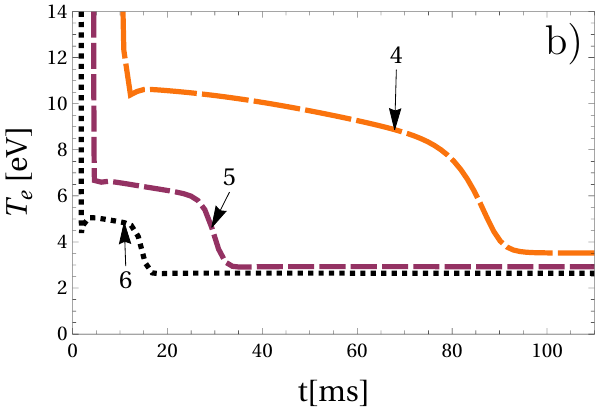}
    \includegraphics[width=0.325\textwidth]{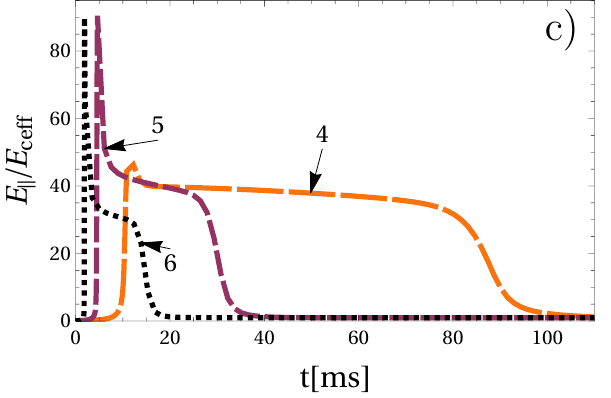}
    \caption{\label{fig:Allcases456}  Time evolution of quantities of interest for the low $\ndi$ representative cases: C4--C6. Line color darkens and dashing shortens with increasing case number, and case numbers are indicated with callouts. a) Runaway electron current. b) Electron temperature at mid-radius. c) Electric field normalized to critical electric field at mid-radius. (Note the longer time range plotted in panel a)}    
\end{figure}

The cases at $\ndi=10^{20}\,\rm m^{-3}$ -- C4 to C6 -- are not affected by hydrogen recombination as their temperature never drops below $2\,\rm eV$. They reach much higher values of $\Epa/\Eceff$ than the high $\ndi$ cases, as they have low $\Eceff$; compare Figs.~\ref{fig:Allcases123}c  and \ref{fig:Allcases456}c. Their RE conversion timing and magnitude well correlates with when the peak of $\Epa /\Eceff$ is reached, and its magnitude. This, in turn depends on the first equilibrium temperature reached, varying between approximately $5$ and $11\,\rm eV$, see Fig.~\ref{fig:Allcases456}b. This temperature decreases monotonically with increasing injected neon quantity, while the magnitude of the final RE current increases, and the time of RE conversion shifts earlier. Once the conversion is complete, the temperature falls further into the $2$--$4\,\rm eV$ range. Note that at these low $\ndi$ cases the dissipation rate of the RE current in the RE plateau is negligible during the simulation, due to the lower $\Eceff$ values.   

Only in C4 does $t_{\rm CQ}$ fall in the acceptable range, in the other two cases it is too short due to the early RE conversion. The reason for the non-monotonic dependence of  $\mathcal{L}$ with increasing $\nnei$, i.e.~it is higher for C5 than for C6, is caused by the reduction in the transported heat loss fraction from the $70$--$80\%$ range to $23\%$ (that is still not acceptably low though).

\subsection{Radially varying material injection}
\label{sec:radvar}

Next, we relax the assumption of spatially homogeneous injection, and allow profile variations with a simple model for the injected densities, where the inward or outward peaking of the profile is set by a single parameter $c_i$ per species $i$
\begin{equation}
 \tilde{n}_{i,\rm inj} \propto 1+\tanh \left[c_i\left(\frac{r}{a}-\frac{1}{2}\right)\right],
\label{eq:profileparam}
\end{equation}
where the tilde indicates that $\tilde{n}_{i,\rm inj}$ is a radially varying quantity. The notation $n_{i,\rm inj}$ is reserved to the scalar parameter that appears in the optimization. The factor multiplying the expression in Eq.~(\ref{eq:profileparam}) is determined such that the total number of injected particles in the plasma is the same as in an injection of a constant density $n_{i,\rm inj}$. Negative/positive values of $c_i$ correspond to densities peaked in the plasma center/edge, and in the optimization we allow values in the $[-10,\, 10]$ range. 

\begin{figure}
    \centering
    \includegraphics[width=0.32\textwidth]{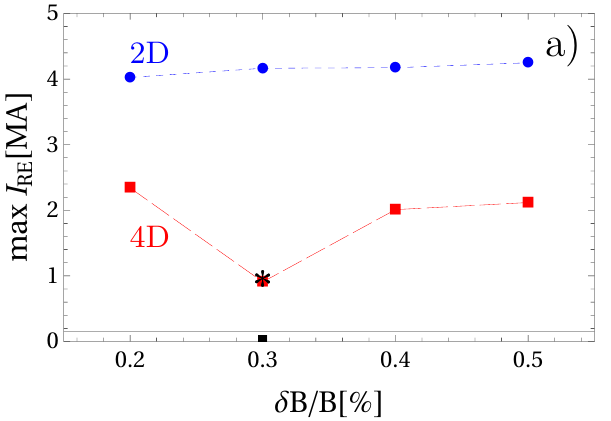}
    \includegraphics[width=0.33\textwidth]{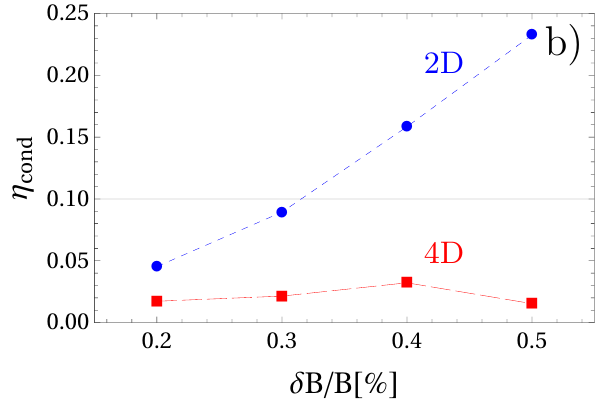}
    \includegraphics[width=0.32\textwidth]{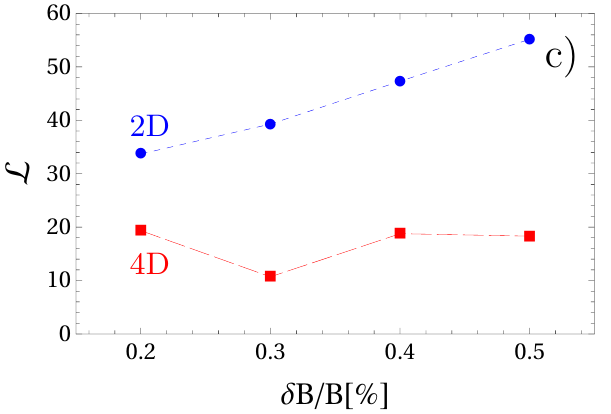}
    \caption{\label{fig:FigsMerit} Variation of (a) the maximum RE current, (b) the transported heat loss fraction, and (c) the corresponding cost function in optimizations, for a range of $\dBB$ values, when optimizing only for injected densities (circle markers, blue short dashed curve) and when including profile variation as well in the optimization (squares, red long dashed). In panels a) and b), below the thin solid line the values are considered tolerable. In panel a) simulations with the parameters corresponding to the 2D optimum at $\dBB=0.3\%$, but without activated sources is indicated with a black rectangle marker, and a simulation with a reduced wall radius of $2.15\,\rm m$ is shown with a black asterisk.}    
\end{figure}

Figure~\ref{fig:FigsMerit} shows the $I_{\rm RE}^{\rm max}$  and $\eta_{\rm cond}$ figures of merit, along with the cost function $\mathcal{L}$ at the optima found for different $\dBB$ values, when radially constant injection is employed (dotted line, referred to as 2D) and when profile variation is allowed (dashed, 4D). In the latter case, the additional degrees of freedom allow us to find optima with better properties. Since in all cases the remaining ohmic current is much smaller than $I_{\rm RE}^{\rm max}$ (in the $300$-$400\,\rm kA$ range), and $t_{\rm CQ}$ is also in the tolerable range, $\mathcal{L}$ is dominated by the two figures of merit plotted. In none of the cases considered is $I_{\rm RE}^{\rm max}$ tolerably small; it is around $4\,\rm MA$ independently of $\dBB$ in the 2D optimization, and it reduces almost by a factor of 2 in 4D (without any clear trend with $\dBB$), as seen in Fig.~\ref{fig:FigsMerit}a.

There are two main reasons for obtaining such high values even in the optimal cases. We consider D-T plasmas, and we include RE seed sources relevant for activated operation, tritium decay and Compton scattering of $\gamma$ photons, in addition to Dreicer and hot-tail RE generation. The tritium decay and Compton sources can provide a significant RE seed even after the TQ, during which the transport due to magnetic perturbations decimates the initial hot-tail and Dreicer seed population. This circumstance also explains the weak sensitivity of $I_{\rm RE}^{\rm max}$ to $\dBB$ in the 2D simulations. A simulation identical to the 2D optimum at $\dBB=0.3\%$, but without activated seed sources (i.e.~only Dreicer, hot-tail and avalanche sources active) yields a negligibly small $I_{\rm RE}^{\rm max}=4.1\,\rm kA$ instead of $4.2 \,\rm MA$. 

A similarly important factor is the realistic radius of the conducting wall, which is chosen to match the energy in the poloidal magnetic field due to the plasma current within the conducting wall to that observed in JOREK simulations. If in the 2D optimum at $\dBB=0.3\%$ we reduce the wall radius from $2.833\,\rm m$ to $2.15\,\rm m$, which was used in previous work, e.g.~by \cite{Vallhagen2020}, the RE current reduces to the -- non-negligible, but still significantly lower -- value of $1\,\rm MA$.   

The fraction of transported heat losses, shown in Fig.~\ref{fig:FigsMerit}b, increases strongly with $\dBB$ in the 2D cases, which is not surprising, since the heat transport during the TQ is then increasing, while the radiated losses are not directly impacted by $\dBB$. However, when profile variation is allowed $\eta_{\rm cond}$ is almost independent of $\dBB$; the reason for this will be explained in relation to Fig.~\ref{fig:4Dvs4D}.    

\subsection{Sensitivity of the optima}
\label{sec:sensitivity}
\begin{figure}
    \centering
    \includegraphics[width=0.5\textwidth]{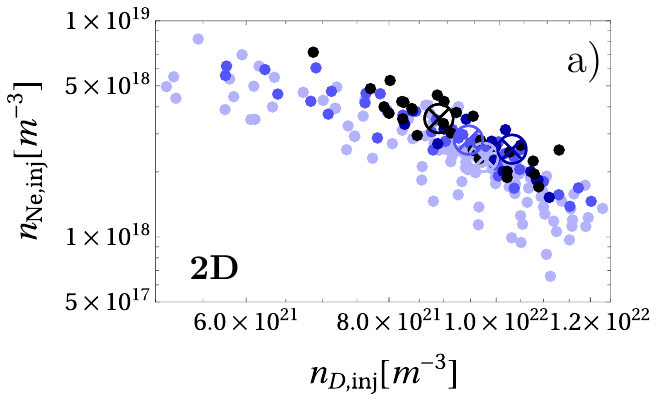}
    \includegraphics[width=0.4715\textwidth]{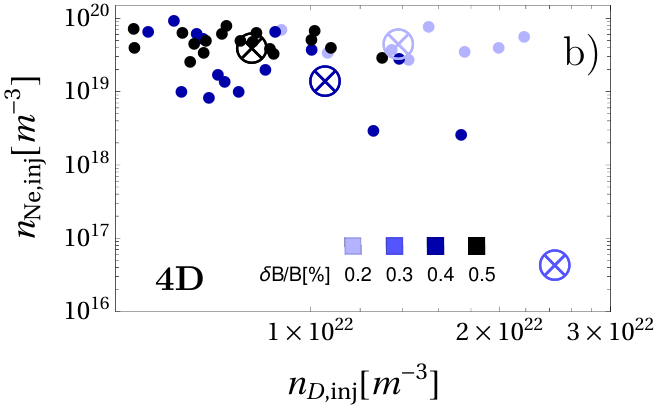}
    \\
    \includegraphics[width=0.483\textwidth]{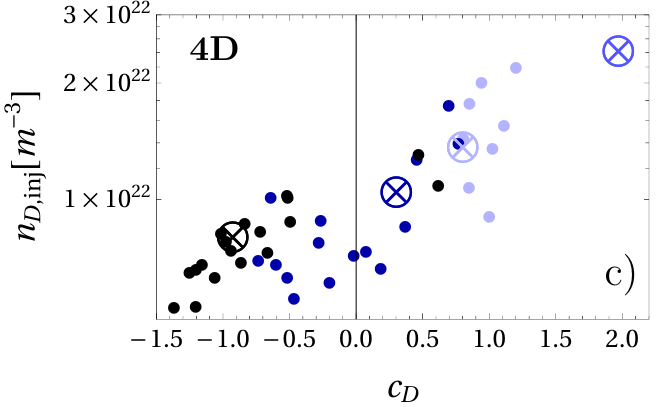}
    \includegraphics[width=0.4485\textwidth]{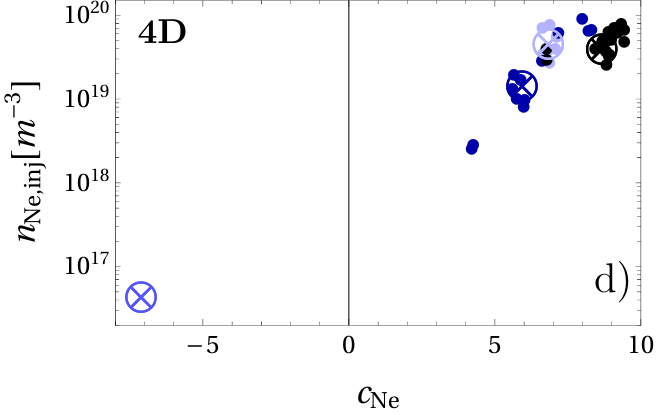}
    \caption{\label{fig:Clouds} Scatter plot of input parameters for samples with the lowest $\mathcal{L}$ values in each optimization case. When (a) optimizing only for injected densities (2D) they represent an additional $10\%$ range above the optimum, and when (b-d) including profile variation as well in the optimization (4D), they represent a $25\%$ range. Darkening color indicates increasing value of 
    $\dBB$, as given in panel b),  and the optima are indicated by $\otimes$ markers. (a-b) Concentration space, (c-d) correlating concentration with profile parameter of an injected species. Note that in the 4D, $\dBB=0.3\%$ case there is no sample within the $25\%$ range above the optimum.}    
\end{figure}

 To gauge the sensitivity of the optima to the input parameters, we investigate the regions occupied by samples within some range of $\mathcal{L}$ above the optimal values. The location of the optima in the optimization space is marked in Fig.~\ref{fig:Clouds} ($\otimes$ markers). In the 2D optimization study we also scatter-plot all samples in the $10\%$ vicinity of the optimum, Fig.~\ref{fig:Clouds}a; this is such a narrow range in $\mathcal{L}$, that any point in this point cloud can be considered equally well performing as the optimum itself. In the 4D optimization study we show points in the  $25\%$ vicinity of the optima, Fig.~\ref{fig:Clouds}b-d. As the total number of samples is the same in both the 2D and 4D optimization studies, the higher dimensionality in 4D implies a sparser exploration in the vicinity of the optimum compared to 2D; hence the lower number of points in spite of the wider relative range included. 

First, considering the 2D optimization, Fig.~\ref{fig:Clouds}a, we find that the relative extent of the point clouds is significantly larger in the $\nnei$ direction, than in the $\ndi$ direction; for instance in the $\dBB=0.2\%$ case $\nnei$ spans more than an order of magnitude, while $\ndi$ spans only a bit more than a factor of two. In practice it translates to the need of a higher precision concerning the injected amount of deuterium than that of neon. The negative correlation between $\ndi$ and $\nnei$ seen from the arrangement of the point cloud indicates that there are similarities in the effects of these two injected species. These features are also reflected in the favourable valleys (blue tone regions) seen in Fig.~\ref{fig:2x2}. The favourable parameter range indicated by the point clouds shrinks with $\dBB$. Note that the region covered by the optima at different $\dBB$ values  is even smaller than the smallest (black) point cloud; thus we should not read much into how the actual location of the optima varies with  $\dBB$.

\begin{table}
  \begin{center}
  \begin{tabular}{ccccc}
      $\dBB$ & $n_{\rm D}(0)[10^{20}m^{-3}]$ & $n_{\rm D}(a)[10^{20}m^{-3}]$ & $n_{\rm Ne}(0)[10^{18}m^{-3}]$ & $n_{\rm Ne}(a)[10^{18}m^{-3}]$  \\[3pt]
 0.2\% & 80 & 160 & 0.14 & 62 \\ 
 0.3\% & 55 & 320 & 0.19 & $3\cdot 10^{-4}$ \\
 0.4\% & 87 & 110 & 0.10 & 20 \\
 0.5\% & 140 & 60 & 0.02 & 54  \\
  \end{tabular}
  \caption{\label{tab:densities} Total hydrogenic (including the background) and neon densities at the plasma center ($r=0$) and at the edge ($r=a$) in the 4D optimization in the various $\dBB$ cases.}
  \end{center}
\end{table}

In the 4D optimization, the resulting point clouds are more scattered, when projected into the $\ndi$--$\nnei$ subspace, see Fig.~\ref{fig:Clouds}b. If anything, there is still a weak anti-correlation between the injected quantities, but the poor statistics makes it less clear. Similarly to the 2D optimization,  the range covered in $\ndi$ is smaller than that in $\nnei$. We can also see that there are no cases within a relative range of $25\%$ of the optimum for $\dBB=0.3\%$. In addition, the optimum itself appears far in the parameter space from the other three overlapping clouds. Namely, it appears at the highest $\ndi$ and lowest $\nnei$ values. We omit this outlier case in the following discussion, but will return to it at the end of this section.

The point clouds occupy the relatively narrow $c_{\rm D}\in [-1.5,\,1.2]$ range, as seen in Fig.~\ref{fig:Clouds}c, corresponding to modest profile variation.  We find a positive correlation between $\ndi$ and $c_{\rm D}$. It means that higher injected content corresponds to more edge-localized peaking. In particular, the injected densities at the plasma center occupy a narrower range than at the edge (see Table~\ref{tab:densities}); apparently, the deuterium density value at the edge is less important.  We also observe that lower $\dBB$ corresponds to higher $c_{\rm D}$ and $\ndi$ values.  

For the injected neon profiles, a strong outward peaking is preferred, with values of $c_{\rm Ne}\in [5,\,10]$, as seen in Fig.~\ref{fig:Clouds}d. The total injected quantities are typically higher than those in the 2D optimization, covering mostly the $\nnei\in [10^{19},\,10^{20}]\rm m^{-3}$ range -- an order of magnitude higher than in 2D. It is interesting to note that, similarly to deuterium, there is a positive correlation between $c_{\rm Ne}$ and $\nnei$.    

\begin{figure}
    \centering
    \includegraphics[width=0.33\textwidth]{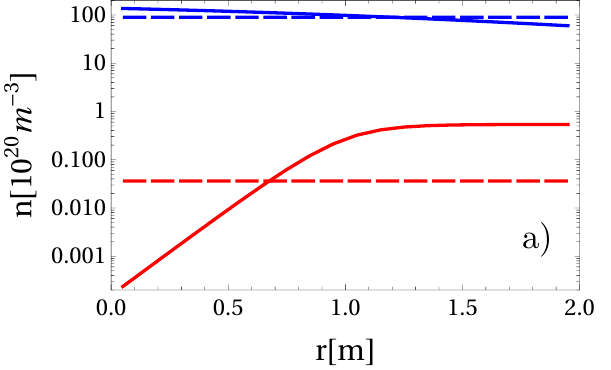}
    \includegraphics[width=0.315\textwidth]{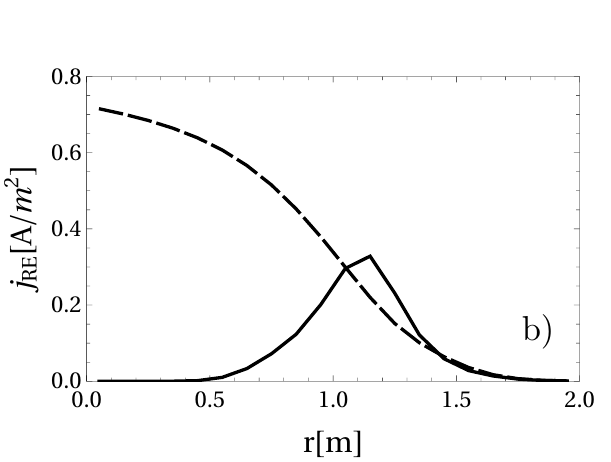}
    \includegraphics[width=0.325\textwidth]{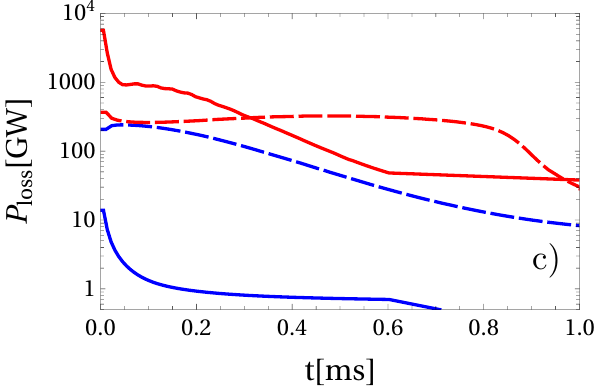}
    \caption{\label{fig:4Dvs4D} Comparison of the optimal cases in the 2D (dashed curves) and the 4D (solid curves) optimization for $\dBB=0.5\%$. a) Radial total hydrogenic density, $n_{\rm D+T+D,inj}$ (blue), and neon density, $n_{\rm Ne}$ (red). b) RE current density profiles taken at the time point when the total RE current takes its maximum, $t=42\,\rm ms$ ($50\,\rm ms$) in the 2D (4D) case. c) Time evolution of the heat loss power in the first millisecond, when most of the thermal energy is lost from the plasmas (note the log-scale). Blue curves represent the transported heat losses, red curves are the radiated losses. }   
\end{figure}

To understand why the optima in the 4D optimization perform better than those of 2D, we compare the respective $\dBB=0.5\%$ cases, where the figures of merit are most disparate. The hydrogenic (blue curves) and neon (red) density profiles of the 2D (dashed curves) and 4D (solid) optima, are shown in Fig.~\ref{fig:4Dvs4D}a. For deuterium, the 4D optimization finds a moderate inward peaking ($c_{\rm D}=-0.92$), while the neon profile is strongly peaked at the edge ($c_{\rm Ne}=8.67$), covering a density range over three orders of magnitude. 

The neon content has two major effects on our figures of merit. An increasing neon concentration corresponds to a lower quasi-equilibrium temperature during the RE conversion, typically leading to higher final RE currents. This is the same trend that we have witnessed moving from C4 to C6. At the same time, a higher neon concentration can help increasing the radiated fraction of heat losses (this was also clear when comparing C5 to C6). However, the neon concentration affects the final RE current most strongly where the RE growth is strongest. This happens to be the plasma core in the parameter region of interest, without a radial variation of the neon density. In addition, to achieve a low $\eta_{\rm cond}$ value it is sufficient to have enough radiating impurities in the edge. Both requirements can be satisfied by an outward peaking neon concentration, which is indeed what the 4D optima tend to develop.        

We find that the 2D optimum produces a centrally peaked RE current, as seen in Fig.~\ref{fig:4Dvs4D}b (dashed curve), while the 4D optimum has a RE profile peaked off-axis (solid curve), as expected for the low core concentration of neon. We note, that in this case the runaway and the centrally peaked ohmic currents decay together after the RE current reached its maximum, and only towards the end of the simulation ($\approx 140\,\rm ms$) does the total current profile become truly hollow\footnote{The magnetohydrodynamic stability of the current density is not monitored in the \dream{} simulations; a hollow current profile might well be unstable to macroscopic plasma instabilities; this aspect of the simulated current evolution is outside the scope of this study.}.   

The time evolution of the volume-integrated heat losses is shown in Fig.~\ref{fig:4Dvs4D}c in the first millisecond. This is when the vast majority of the thermal energy content of the plasma is lost, while the fraction of magnetic-to-thermal energy conversion is still negligible. Again, the dashed curves correspond to the 2D optimum; in this case the transported loss (dashed blue) reaches comparable values to the radiated losses (dashed red). The entire energy loss process varies relatively smoothly over the plotted timescale. In contrast, in the 4D optimum case the transported heat losses (solid blue) are approximately two orders of magnitude lower than the radiated losses (solid red), and both of these channels have a strong peak at $t=0$, related to the ionization and equilibration of the injected material. 

After having discussed the representative behavior at the optima we return to the analysis of the outlier case, the 4D optimum at $\dBB=0.3\%$. In this case the injected neon density is roughly three orders of magnitude lower than in the other three cases, and as such, it exhibits reheating following the TQ in the plasma center. This reheated region supports a relatively slowly decaying ohmic current, hence the CQ time is on the long side $t_{\rm CQ}=123\,\rm ms$ (while still tolerable). The slowly decaying ohmic current and the high value of the effective critical electric field $\Eceff$, owing to the high $\ndi$, lead to that the RE growth stops just before the RE current grows to macroscopic values. The strong dilution is able to rapidly reduce the temperature to sufficiently low values at the edge, that even in the presence of a low neon content the cooling can continue to $\approx 1 \,\rm eV$. As then most of the heat transported to the edge is radiated away by the recombined deuterium, the resulting transported heat loss also remains small in this case.  This is a fragile case nevertheless; indeed there is no sample within $25\%$ of the $\mathcal{L}$ value reached by this optimum. Some parameter combinations in the vicinity of this optimum yield a behavior reminiscent of C3, with an extremely rapid RE conversion and then a strongly decaying RE current. Thus, even though this optimum performs better than the other three cases in 4D, it should not be targeted in a experiment, due to the lack of robustness. 

Finally, we comment on the  numerical efficiency of the Bayesian approach.  We estimate that to achieve a similar level of resolution in the regions that contain samples within $25\%$ of the optima would require more than $12\,000$ points in 2D and $800\,000$ points in 4D, should we decide to use equidistant scans over the entire search domains. These estimates are based on the average minimum distance between samples (in the search space mapped to the unit hyper-cube). As a reference, we use only 420 samples in both the 2D and the 4D optimizations. In uninteresting regions with high cost function values the resolution is much lower.

The Bayesian results can be confirmed with calculations on a uniform grid. In a detailed study of a similar problem presented by \citet{Bergstrom2022}, it was shown that the mean function obtained by the Gaussian process regression accurately recovered the cost function calculated on a uniform grid in the vicinity of the optimum, and showed a good agreement even in regions with high cost values. In terms of finding the global optimum the Bayesian method outperformed Powell's method \citep{powell64}.

\section{Discussion and conclusions}
\label{sec:conclusions}
We have used Bayesian optimization to find optimal parameters characterizing massive material injection. This is a multi-objective problem where the cost function we aim to minimize accounts for the maximum RE current, the transported heat loss fraction, the CQ time, and the final ohmic current.  Bayesian optimization is well suited for this problem, as it is a computationally efficient method for finding global optima, providing also uncertainty quantification. In the disruption context, it has also been used recently for validation of simulations of a CQ in a JET plasma discharge with an argon induced disruption \citep{Jarvinen_2022}.

 We find that  even in the optimal case, RE currents of several megaampere are predicted. Magnetic perturbations strongly affect the RE dynamics through inducing transport losses of heat and seed REs. Then the optimization is, to a large degree, searching for a balance between sufficiently low transported heat loss -- typically favoring large injected impurity quantities and low magnetic perturbation amplitudes -- and tolerable final RE current -- favoring the opposite conditions. The importance of such a balance has previously been pointed out by \citet{Svenningsson2021}.  In each optimization  we kept the normalized magnetic perturbation level constant, in the range $0.2$-$0.5\%$. This range of magnetic perturbation levels  is motivated by MHD simulations, We note that  higher values are also reached in some recent studies \citep{Nardon_2021,KonstaNF2020}, which, based on the trends we observe in Fig.~\ref{fig:FigsMerit}, is not expected to have significant effect on the final RE current, while it would impact the transported heat loss fraction negatively. 
 
 The optimum is generally found at a rather high injected deuterium density $n_{\rm inj,D}\approx  10^{22}\,\rm m^{-3}$, while at a lower neon density $n_{\rm inj,Ne}\approx 3\times 10^{18}\,\rm m^{-3}$. The sensitivity of the optimum to an inaccuracy of the injected deuterium quantity is much stronger than that of the injected neon. The strong sensitivity to the deuterium quantity is due to the possibility of extremely rapid cooling through dilution and subsequent radiation at sufficiently high deuterium densities, which leads to an effective seed generation. In addition, deuterium recombination steeply increases above a certain deuterium density, allowing the already large seed to avalanche more effectively. We also find, that neon deposited at the edge is advantageous, where it can produce sufficient radiative heat losses, without making the avalanche RE generation problem more severe, for which the conditions are typically more favorable in the core. Whether an outward peaking impurity density can be sustained long enough to see these benefits can only be answered using higher fidelity simulations. In this sense, our 4D optimization results can be considered as optimistic bounds.
 
 We point out the importance of choosing the wall radius carefully, as it determines the magnetic energy reservoir for RE generation; a tightly fitted conducting wall may lead to too optimistic results concerning the maximum RE current (yielding $1\,\rm MA$ instead of $4\,\rm MA$ in our example). As we allow for activated RE seed generation mechanisms we cannot find parameter regions where all objectives fall within their respective tolerable ranges; we see however that this may not need to be the case with non-activated seed sources only.   

The megaampere-scale RE currents predicted even in the optimal scenarios is concerning, thus these results should prompt further studies accounting for additional effects that can impact RE current generation.  The most important effects to consider are: 1) magnetohydrodynamic and kinetic instabilities, 2) vertical displacement and the associated interaction of the current-carrying plasma column with the wall, 3) the possibility of magnetic surface re-healing to take place significantly later than the end of the TQ, and 4) the possible disappearance of closed flux surfaces below a finite -- still megaampere-level -- plasma current. In addition, the dynamics of the injection -- which is not resolved here -- has a direct impact on the transported heat fraction, and more generally it may affect the temperature evolution and in turn the RE dynamics (mostly the Dreicer and hot-tail seed generation, and as such, it is expected to be more consequential in non-activated operation). Employing this Bayesian framework for the optimization of the more directly accessible parameters describing the injection (for instance the composition and timing of the injected pellets in shattered pellet injection) is thus a natural next step to pursue. 

The results are quite robust with respect to the choice of the cost function.
The most important trade-off between the various figures of merit appears between achieving a low runaway current and a low transported heat fraction.
For instance, in the $\delta B/B=0.3\%$ 2D case, changing the weight of $\eta_{\rm cond}$ in the cost function by $\pm 10\%$ moves the optimum by $\pm 1.5\%$ in $n_{\rm Ne,inj}$, and by $0.4\%$ in $n_{\rm D,inj}$. These figures are calculated relative to the extent of the $10\%$ neighborhood of the optimum on a logarithmic scale (i.e.,~the size of the corresponding point cloud in Fig.~\ref{fig:Clouds}a). The lower bound of the $10\%$ neighborhood of the optimum changes by $\pm 5\%$, while the other bounds change by $1\%$ or less. The functional form and weight of the various components in the cost function are ultimately chosen by the user.  Currently this arbitrariness of the weights cannot be fully eliminated, partly because of a detailed knowledge about the (monetary) cost of a given value of a figure of merit is lacking, and such estimated figures may never be available. In addition, the current modelling provides too coarse information on the outcome of a given scenario. Indeed, RE beams with the same RE current may cause serious damage, or no detectable effect at all, depending on how the beam is lost to the wall. Recent results indicate that a combination of a low impurity concentration bulk plasma and large-scale magnetohydrodynamic instabilities  may enable termination of megaampere-level RE currents without damage to the wall \citep{Reux2021,PazSoldan_2021}.

\section*{Acknowledgements} 
The authors are grateful to N.~Botta, N.~Smallbone, E.~Berger, S.~Newton, E.~Nardon, J.~Artola and M.~Lehnen for fruitful discussions. 

\section*{Funding} 
This work was supported by the Swedish Research Council (Dnr.~2018-03911 and Dnr.~2022-02862) and in part by the Swiss National Science Foundation. The work has been carried out within the framework of the EUROfusion Consortium, funded by the European Union via the Euratom Research and Training Programme (Grant Agreement No 101052200 — EUROfusion). Views and opinions expressed are however those of the author(s) only and do not necessarily reflect those of the European Union or the European Commission. Neither the European Union nor the European Commission can be held responsible for them.

\section*{Declaration of Interests}
The authors report no conflict of interest.

\appendix

\section{Simulation details}
\label{sec:simulations}

\begin{figure}
    \begin{center}
    \includegraphics[width=0.8\textwidth]{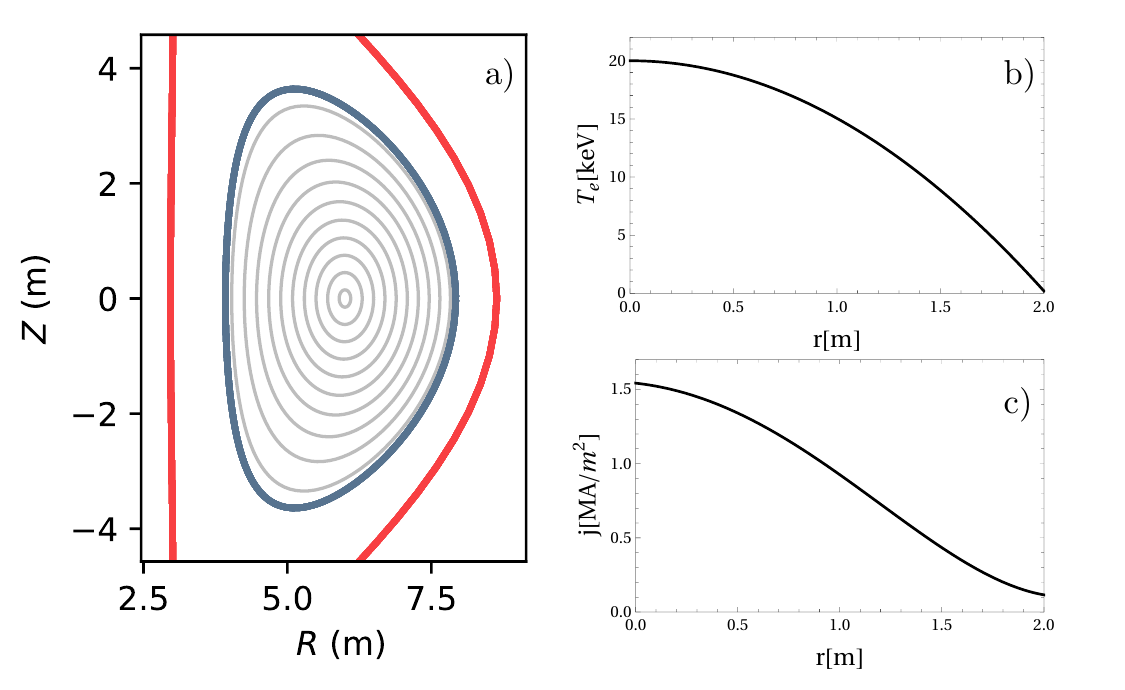}
    \end{center}
    \caption{\label{fig:profiles} a) Magnetic geometry with flux surfaces (gray curves), the outermost modeled flux surface $r=a$ is indicated by the thick blue line, and the effective wall is shown in red. The rest of the panels show initial plasma parameter profiles.  b) Electron temperature. c) Current density. 
    }   
\end{figure}

The magnetic geometry and the initial plasma temperature and current density  profiles are shown in figure~\ref{fig:profiles}a and b-c, respectively. The parallel current density component $j$ is taken at the outboard mid-plane. The magnetic geometry uses a model equilibrium parametrization similar to the Miller equilibrium \citep{Miller98}, with the profiles of elongation, triangularity, Shafranov shift and toroidal magnetic field variation being identical to those shown in Appendix~A of \citep{pusztaiCurrentRelax}.  The on-axis value is $B_0=5.3\,\rm T$. The magnetic equilibrium is not evolved self-consistently in the simulation, instead these shaping parameters, as well as the plasma position, are held fixed throughout the simulation.        

The \dream{} simulations are performed in fluid mode. The Dreicer RE generation rate is calculated using a neural network \citep{Neural}, which takes effects of partial screening into account.
Compton scattering and tritium decay seed sources are accounted for as in \citep{Vallhagen2020}. The hot-tail seed is calculated using the model described in Appendix C.4 in \citep{dreampaper}. The avalanche growth rate accounts for partial screening \citep{Hesslow_2019}. 
Trapping effects are accounted for in the conductivity through the model by \citet{Redl}, and in the avalanche and hot-tail RE generation rates. 

The bulk electron temperature evolution is calculated from the time dependent energy balance throughout the simulation, according to Eq.~(43) in \citep{dreampaper}, accounting for ohmic heating, line and recombination radiation and bremsstrahlung, as well as a radial heat transport. Since the RE population is not resolved in momentum space, the kinetic term -- in Eq.~(44) of \citep{dreampaper} -- describing heating by REs is zero. However, the latter process is approximately accounted for by a term $j_{\rm RE} E_{\rm c}$, with $E_{\rm c}=e^3n_e \ln \Lambda_c/(4\pi \epsilon_0 m_e c^2)$ the critical electric field, $\epsilon_0$ the vacuum permittivity, and $m_e$ the electron mass.   We evolve the temperatures of the ion charge states separately according to Eq.~(45) in \citep{dreampaper} which accounts for collisional heat exchange among various charge states as well as with electrons.  We neglect current density profile flattening \citep{pusztaiCurrentRelax} associated with the flux surface breakup. 

 Opacity effects have been shown to have significant effect  on the post-TQ plasma temperature and indirectly on the avalanche gain \citep{Vallhagen2022}. These effects are taken into account by using ionisation, recombination and radiation rates for the hydrogen isotopes that are
based on the assumption of the plasma being opaque to Lyman radiation.

The simulations use $20$ radial grid cells. During the TQ that takes a few milliseconds the solver uses adaptive time stepping with time steps estimated from the relative change of the free electron density within a time step (referred to as the \emph{ionization-based adaptive time stepping}), with allowed minimum and maximum time steps $10^{-11}\,\rm s$ and $2\times 10^{-6}\,\rm s$. The rest of the $150\,\rm ms$ long simulation uses $2\times 10^{4}$--$2\times 10^{5}$ equidistant time steps as needed for convergence.   

\section{Details of Bayesian optimization}
\label{sec:bayesapp}

After $n$ steps our sample data $\mathcal{D}_n := (X_n, Y_n)$ is a collection of control vectors $X_n=\{\mathbf{x}_i\}$ and the corresponding function outputs $Y_n=\{f(\mathbf{x}_i)\}$ where the function $f$ runs \dream{} to obtain the four objectives and combines them using the cost function $\mathcal{L}$. 
The basic idea of Bayesian optimization is that $f(\mathbf{x})$ is a random variable for each $\mathbf{x}$ and that, given the observations $\mathcal{D}_n$, the joint distribution of all these random variables is a \emph{Gaussian process}.
The corresponding mean and covariance functions are defined as the expected values
\begin{align}
    \mu(\xv)&=\mathbb{E}[f(\xv)],\\
    k(\xv, \xv')&=\mathbb{E}[(f(\xv)-\mu(\xv))(f(\xv')-\mu(\xv'))].
\end{align}
In our case the \dream{} simulation runs are deterministic, which means that the function $\mu$ will exactly coincide with $f$ on the samples observed so far.
In other points the Gaussian process model provides a smooth interpolation of the cost (something we used to visualize the cost function in Fig.~\ref{fig:2x2}).

The covariance between two points is modeled by the Mat\'{e}rn kernel \citep{Matern1986,Stein1999}
\begin{equation}
  k_{\rm M}(\xv,\xv')=\frac{1}{2^{\zeta-1}\Gamma(\zeta)} \left(2 \sqrt{\zeta}|\xv-\xv'|\right)^\zeta K_\zeta \left(2 \sqrt{\zeta}|\xv-\xv'|\right).
  \label{Matern}
\end{equation}
where $\Gamma$ denotes the Gamma function and $K_\zeta$ is the modified Bessel function of the second kind. We use a fixed smoothness parameter of $\zeta=5/2$. The distance between two points in the D-dimensional parameter space is calculated as $|\xv-\xv'|=\sum_{i=1}^D (x_i-x_i')^2/\theta_i^2$, with  the correlation length parameters $\theta_i$ (which are updated after each new sampling to maximize the marginal likelihood of $\mathcal{D}_n$).

We use the \textit{expected improvement} $\mathrm{EI}_n(\xv)$ acquisition function to find the most promising next point to sample. The following thought experiment \citep{Frazier2018} illustrates this acquisition strategy.
Let $f_n^*$ be the minimal value of $f$ based on the current sample, and let $\xv_n^*$ be the corresponding input.
If the optimization procedure is terminated at this sample size, $\xv_n^*$ would be returned as the best estimate of the actual optimum location~$\xv^*$.
Suppose that an additional evaluation is to be performed at any point $\xv$ yielding $f(\xv)$.
After this, the minimal observed value of $f$ is either $f(\xv)$ if $f(\xv)<f_n^*$ or remain to be $f_n^*$ otherwise.
We might define the \emph{improvement} we gain by performing this additional  evaluation to be $f_n^*-f(\xv)$ in the former case -- the amount we could decrease the best value found so far -- and $0$ in the latter.
We aim to maximize this improvement, while $f(\xv)$ is, as of yet, still unknown.
Instead, he next sample location is chosen to maximize the expectation value of the improvement, given the information at hand, that is
\begin{align}
    \label{eq:maximizer}
    \xv_{n+1}=\mathrm{argmax}\Big\{\mathbb{E}_n\big[\mathrm{max}(0, f_n^*-f(\xv))\big]\Big\}=\mathrm{argmax}\big\{\mathrm{EI}_n(\xv)\big\},
\end{align}
where $\mathbb{E}_n[\cdot]$ should be understood as the expectation under the posterior distribution, given the previously evaluated $\mathcal{D}_n$.

\bibliographystyle{jpp}
\bibliography{bibliography}

\end{document}